\documentclass{optica-article}

\journal{opticajournal} 

\articletype{Research Article}
\usepackage{algorithm,algorithmic}
\usepackage{lineno}
\usepackage{booktabs}

\begin{document}

\title{Parallel overlapping-domain decomposition FDFD for large-scale complex nanostructures modeling}

\author{Zhanwen Wang,\authormark{1} Chengnian Huang,\authormark{1} Wangtao Lu,\authormark{2} Yuntian Chen,\authormark{3,4} and Wei E. I. Sha,\authormark{1,5}}

\address{\authormark{1} the College of Information Science and Electronic Engineering, Zhejiang University, Hangzhou 310027, China\\
\authormark{2} the School of Mathematical Sciences, Zhejiang University, Hangzhou 310027,China\\
\authormark{3}School of Optical and Electronic Information, Huazhong University of Science and Technology, Wuhan 430074, China\\
}
\email{\authormark{4}yuntian@hust.edu.cn}
\email{\authormark{5}weisha@zju.edu.cn} 


\begin{abstract*} 
 The increasing complexity and scale of photonic and electromagnetic devices demand efficient and accurate numerical solvers. In this work, we develop a parallel overlapping domain decomposition method (DDM) based on the finite-difference frequency-domain (FDFD) formulation to model the electromagnetic response of large-scale complex nanostructures. The global computational domain is partitioned into multiple overlapping subdomains terminated with perfectly matched layers (PMLs), enabling seamless source transfer between adjacent subdomains. A multi-frontal preconditioner is employed to accelerate the iterative solution process, while an OpenMP-based parallel implementation ensures high scalability. Several numerical examples are provided to validate the efficiency and accuracy of the proposed algorithm. The results demonstrate excellent agreement with analytical and commercial COMSOL solutions. Notably, the method achieves up to an order of magnitude reduction in computation time, highlighting its potential as a powerful tool for large-scale photonic and electromagnetic modeling.

\end{abstract*}

\section{Introduction}
Recent advances in electromagnetic and photonic technologies have created a pressing demand for numerical solvers capable of handling systems with extremely large scales and intricate geometries. Emerging devices—including large-scale photonic integrated circuits (PICs) \cite{He2022,Ning2024}, metasurface \cite{Huang2024,Malevich2025}, and massive antenna arrays \cite{Taygur2019}—often involve complex geometries, multi-scale interactions, and millions to billions of degrees of freedom (DoFs). Accurate and efficient modeling of such systems is crucial for the development of applications ranging from high-speed optical interconnects to next-generation wireless communications. However, traditional full-wave solvers, such as the finite-difference time-domain (FDTD) \cite{Sullivan2013} and the finite-element method (FEM) \cite{Jin2015}, face prohibitive memory and time costs when tackling large electrical sizes, multiscale features, and strong near-field coupling. For instance, large-scale photonic crystal slabs \cite{Minkov2020}, diffractive optical waveguides \cite{Mukawa2009,Kramer2023,Low2022} and metasurface-based flat lenses \cite{Arbabi2016,Su2018,Ding2022} often require simultaneous resolution of subwavelength features and global field distributions, leading to extremely large sparse linear systems. The increasing computational burden is becoming a major barrier for the rapid design and optimization of next-generation photonic and electromagnetic devices.

To address these challenges, several efficient strategies have been proposed. Locally periodic approximation (LPA) simplifies metasurface and metalens design by neglecting long-range coupling \cite{Pestourie2018,Skarda2022,Zu2025,Li2022}. Hybrid wave–ray strategies combine rigorous full wave solvers (e.g., FDTD, Fourier optics, FEM) with global ray tracing (RT) propagation, enabling scalable modeling of diffractive waveguide arrays, multilayer diffractive optics, and large-area diffraction-coupled systems\cite{Xiao2025,Wyrowski2011,Wang2024,Lokar2019a}. This approach preserves key wave phenomena such as diffraction, coupling, and near-field scattering, while leveraging the computational efficiency of ray methods for large-scale propagation. Despite their efficiency, these approximations inevitably sacrifice accuracy in strongly coupled or non-periodic scenarios.

Domain decomposition method (DDM) provides a powerful “divide-and-conquer” framework for large-scale electromagnetic simulations, in which the computational domain is partitioned into smaller subdomains, which can be solved independently and in parallel. Over the years, techniques such as "cement" element method \cite{Lee2005,Vouvakis2006}, finite-element tearing and interconnecting (FETI) algorithm \cite{Xue2012,Gao2016} and  higher-order transmission conditions \cite{Peng2011,Jia2019} have been successfully applied to the analysis of massive antenna arrays, electromagnetic bandgap structures and frequency-selective surfaces(FSS) and so on \cite{Zhao2016,Zhao2020,Wang2017,MacKie2018}. More recently, Zhiming Chen introduced the source transfer domain decomposition method (STDDM) for layered media \cite{Chen2013,Xiang2013}, in which sources are transferred layer by layer, enabling efficient solutions with perfectly matched layers (PML). Parallel STDDM has shown excellent performance in solving velocity models and multilayer propagation problems \cite{Leng2015,Leng2022,Dai2022}, but their applications have been limited to relatively simple geometries and excitations conditions.

In this paper, we extend the STDDM framework to the large-scale complex nanostructures using the finite-difference frequency-domain (FDFD) method. Unlike previous studies that primarily focus on mathematical convergence analysis or relatively simple propagation models, our approach supports arbitrary geometries and diverse excitation types, including scattering sources and mode sources, thus greatly broadening the applicability of STDDM in practical electromagnetic analysis. To further enhance computational efficiency, we introduce multi-frontal preconditioner \cite{Xia2010} that reduces each iteration to an efficient matrix–vector multiplication, based on the fact that only the source terms need to be modified during iterations. Moreover, we implement the proposed method with an efficient OpenMP-based parallelization strategy and systematically analyze the impact of subdomain partitioning on parallel efficiency, offering practical guidelines for large-scale simulations.

This paper is organized as follows: we introduce the overall framework and basic theory of the proposed method in Section 2. In Section 3, the efficiency and superiority of the proposed method are demonstrated through two numerical examples: the scattering of a dielectric cylinder and a large-scale trapezoidal-shaped topological optical waveguide. Finally, the conclusion is reached in section 4.

\section{Basic theory}
\subsection{Finite-difference equation}
The FDFD method is a well-established numerical technique for solving Maxwell’s equations in the frequency domain. By discretizing the computational domain using a structured grid, the partial differential equations governing electromagnetic wave propagation are transformed into large sparse linear systems. The FDFD approach is especially attractive for modeling complex materials and geometries, as it directly accommodates inhomogeneous and anisotropic media.

In two-dimensional transverse magnetic (TM) polarization, the wave equation of the total electric field $E_z^t$ is given by:
\begin{equation}
\nabla \cdot \left( \frac{1}{\mu_r} \nabla E_z^t \right) + k_0^2 \varepsilon_r E_z^t = 0,
\label{equ_total}
\end{equation}
where $\varepsilon_r$ and $\mu_r$ represent the relative electric permittivity and magnetic permeability, respectively. In this work, we assume that $\mu_r=1$. $k_0=\frac{2\pi}{\lambda}$ is the free-space wavenumber, where $\lambda$ is the wavelength. 

\begin{figure}[htbp]
\centering
    \includegraphics[width=0.8\linewidth]{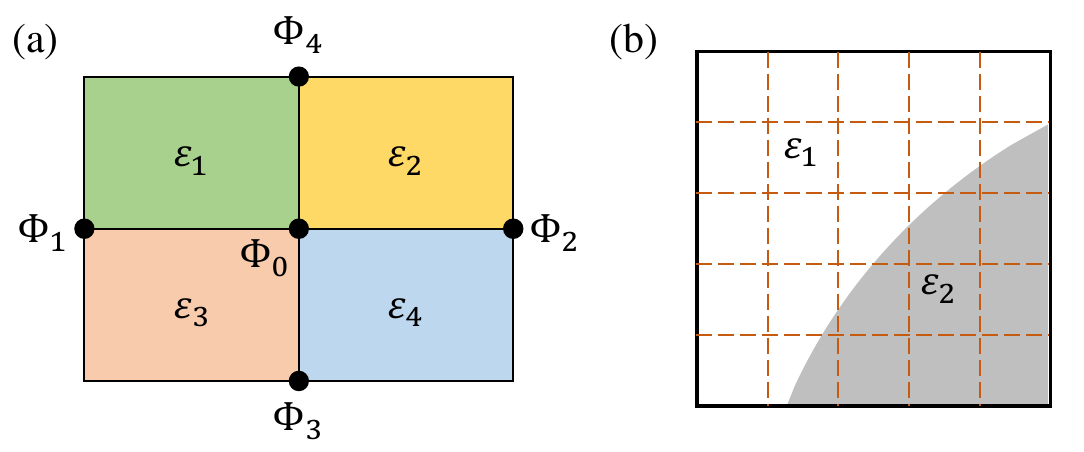}
\caption{(a) The five-point stencil for the FDFD method, where $\Phi_0$ represents the central node and $\Phi_1$ to $\Phi_4$ denote adjacent nodes. $\varepsilon$ is the relative permittivity in the discretized region. (b) The subcell method for material interfaces. The relative permittivity $\varepsilon_1$ and $\varepsilon_2$ are averaged across fine subcells to accurately resolve boundary effects.}
\label{fig_FD}
\end{figure}

As shown in Fig. \ref{fig_FD}(a), using the notations of $\Phi_0=E_z^t(i,j)$, $\Phi_1=E_z^t(i-1,j)$, $\Phi_2=E_z^t(i+1,j)$, $\Phi_3=E_z^t(i,j-1)$, $\Phi_4=E_z^t(i,j+1)$, the discretized FDFD forms for the wave equation Eq. (\ref{equ_total}) is of the form \cite{Qiao2011}
\begin{equation}
-2\left(\frac{1}{\Delta_x^2}+\frac{1}{\Delta_y^2}\right)\Phi_0+k_0 ^2\bar{\varepsilon} \Phi_0+\frac{\Phi_{1}+\Phi_{2}}{ \Delta_{x}^{2}}+\frac{\Phi_{3}+\Phi_{4}}{\Delta_{y}^{2}}=0,
\label{equ_fd}
\end{equation}
\begin{equation}
\bar{\varepsilon}=\frac{\varepsilon_1+\varepsilon_2+\varepsilon_3+\varepsilon_4}{4},
\end{equation}
where $\Delta_x$ and $\Delta_y$ are the spatial step along the $x$-direction and $y$-direction, respectively. This finite-difference equation is enforced at each interior grid node $(i,j)$, leading to a large sparse linear system, which is solved for the field $E_z^t$ under proper boundary conditions and source:
\begin{equation}
\mathbf{A}\mathbf{e}=\mathbf{b},
\label{equ_sys}
\end{equation}
where $\mathbf{A}$ is a sparse matrix with the structure of Eq. (\ref{equ_fd}) on uniform grids, $\mathbf{e}$ is the unknown total field vector and $\mathbf{b}$ represents the discrete source term.

For scattering problems, the total field $E_z^t$ is  decomposed into the incident field $E_z^{inc}$ and the scattered field $E_z^{s}$:
\begin{equation}
E_z^t=E_z^{inc}+E_z^{s},
\end{equation}
The incident field $E_z^{inc}$ satisfies the wave equation in the background medium (assuming air):
\begin{equation}
\nabla^2 E^{inc}_z + k_0^2 E^{inc}_z = 0,
\label{equ_inc}
\end{equation}
Subtracting Eq. (\ref{equ_inc}) from the Eq. (\ref{equ_total}) yields a differential equation for the scattered field \cite{Sha2010}:
\begin{equation}
\nabla^2 E^{s}_z + k_0^2\varepsilon_r E^{s}_z = -k_0^2 (\varepsilon_r-1) E_z^{inc}.
\end{equation}

This scattered field equation can also be discretized using the same finite-difference scheme to get the form of Eq. (\ref{equ_sys}) where the right-hand side contains the equivalent source term based on the incident field.

For the cells intersected by material interfaces, as illustrated in Fig. \ref{fig_FD}(b), we apply a subcell integration method to accurately compute the effective relative permittivity \cite{Sullivan2013}. Each coarse grid cell is subdivided into fine subcells and then the average dielectric properties is assigned accord
ing to the number of subcells in one medium $N_1$ as well as in the other medium $N_2$. Here, the medium type of each subcell is determined by the location of its center relative to the material interface:
\begin{equation}
\varepsilon_{ave}=\frac{\varepsilon_1N_1+\varepsilon_2N_2}{N_1+N_2},
\end{equation}
This approach significantly improves the accuracy at curved boundaries compared to standard cell-centered assignments.

\subsection{Perfectly matched layer}
To simulate unbounded scattering problems using FDFD, it is essential to truncate the computational domain with an absorbing boundary condition. We employ the PML approach, which introduces complex coordinate stretching \cite{Chew1997a} in Maxwell’s equations to absorb outgoing waves without spurious reflection.

For 2D TM polarization, the PML is implemented via a complex coordinate transformation:
\begin{equation}
\frac{1}{s_{r}(x)}\frac{\partial}{\partial x}\left(\frac{1}{s_{r}(x)}\frac{\partial E_z^s}{\partial x}\right)+\frac{1}{s_{r}(y)}\frac{\partial}{\partial y}\left(\frac{1}{s_{r}(y)}\frac{\partial E_z^s}{\partial y}\right)+k_{0}^{2}\varepsilon_r E_z^s=0,
\label{equ_pml}
\end{equation}
where
\begin{equation}
s_r=
\begin{cases}
1+\frac{j\sigma}{\omega\varepsilon_0},\mathrm{within~PML} \\
1,\mathrm{other} & 
\end{cases},
\end{equation}
where $\varepsilon_0$ is the permittivity of free space, and $\omega$ is the angular frequency of the incident light. The conductivities $\sigma(x)$ and $\sigma(y)$ are typically chosen as a polynomial grading, given by:
\begin{equation}
\sigma_u = \frac{\sigma_{max}}{\Delta} \left( \frac{u - \tfrac{1}{2}}{L} \right)^m, \quad u = 1, 2, \dots, L,
\end{equation}
\begin{equation}
\sigma_{u+0.5} = \frac{\sigma_{max}}{\Delta} \left( \frac{u}{L} \right)^m, \quad u = 0, 1, \dots, L,
\end{equation}
where $\Delta=\Delta_x$ or $\Delta=\Delta_y$ for the PML layers normal to the $x$ axis or $y$ axis. $L$ is the layer number of the PML, and $m$ is the order of the polynomial. In this work, the optimized settings are set to $L = 8$, $m = 2$. The coefficient $\sigma_{max}$ is determined based on the theoretical reflection factor under normal incidence $R(0)$, typically satisfying \cite{Berenger1996}:
\begin{equation}
\sigma_{max}=\frac{(m+1)ln[R(0)]}{2\eta d},
\end{equation}
where $\eta=120\pi$ is the free space wave impedance, and $d$ is the physical depth of the PML region. This ensures that the reflection from the PML is below a specified threshold.

\subsection{PML-based overlapping domain decomposition method}
In large-scale FDFD simulations, solving the global sparse linear system becomes increasingly expensive in terms of memory and computation. To overcome these issues, we develop an overlapping domain decomposition method that leverages the existing PML structure. 

\begin{figure}[htbp]
\centering
    \includegraphics[width=0.9\linewidth]{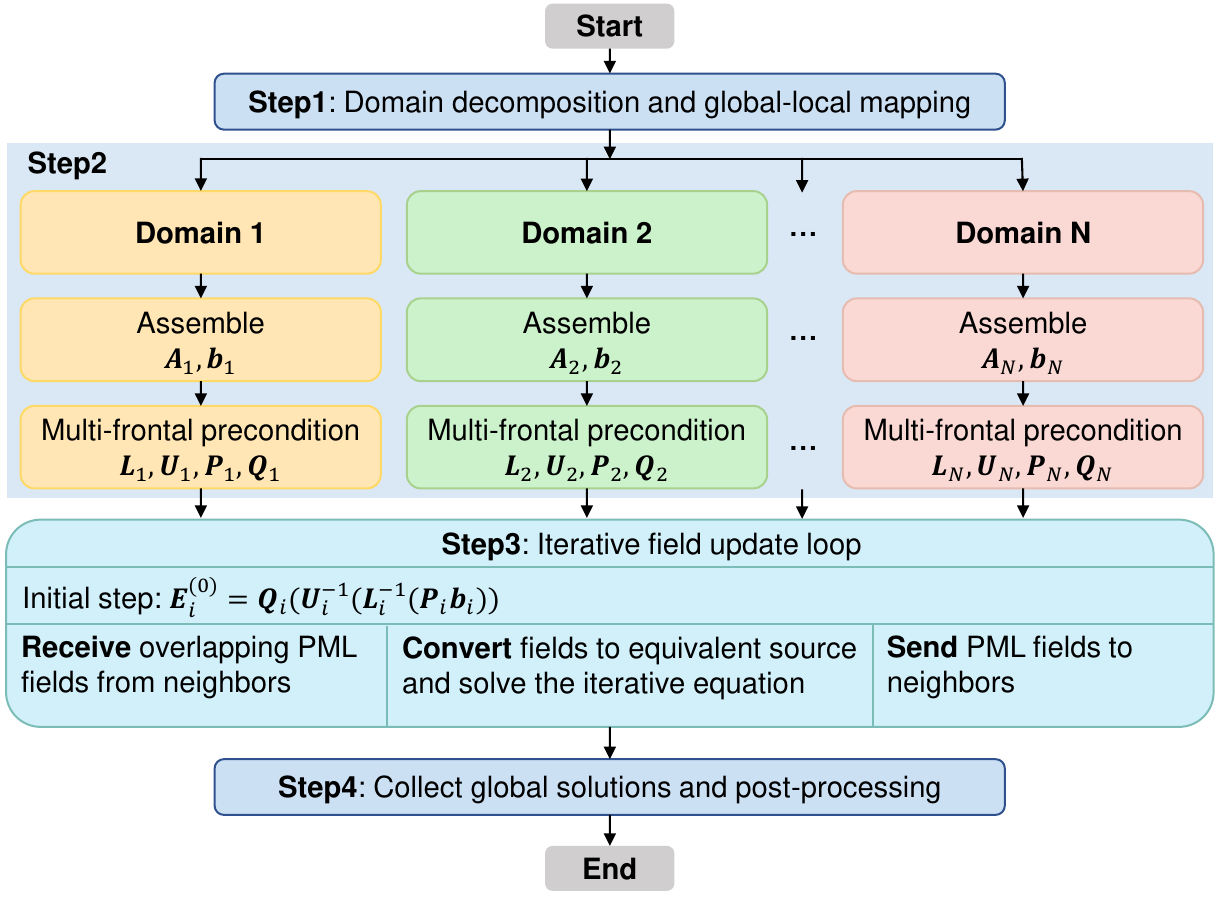}
\caption{The overall framework of the proposed method.}
\label{fig_framework}
\end{figure}

The overall procedure is illustrated in Fig. \ref{fig_framework}. Initially, the simulation data of the model are loaded. The global computational domain is then partitioned into multiple subdomains, each extended with an overlapping PML region that intersects with adjacent subdomains. A global-to-local mapping is constructed to manage data association. Each subdomain independently assembles its local FDFD matrix and performs multi-frontal preconditioning in parallel. In the initial step, the local system is solved using its own right-hand side as Eq. (\ref{equ_sys}). In subsequent iterations, each subdomain receives the field values from the overlapping PML regions of neighboring subdomains, which are used to construct an updated equivalent source. The internal field is then recalculated using the precomputed multi-frontal factors. The updated PML fields are passed back to adjacent subdomains. Once convergence is reached, the global solution is assembled for postprocessing.

\begin{figure}[b!]
\centering
    \includegraphics[width=0.9\linewidth]{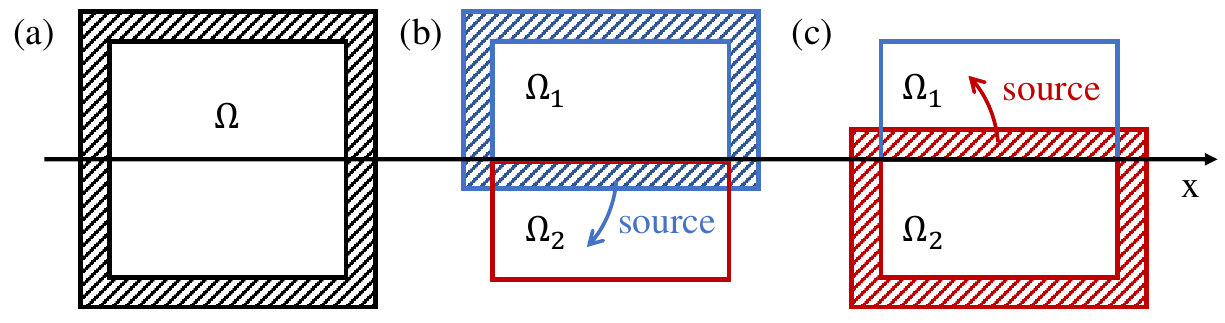}
\caption{Domain decomposition with two subdomains. The hatched area is the PML layer, the overlapping area is the source transfer region. (a) The global computational domain. (b) Subdomain $\Omega_1$ with its local PML region. (c) Subdomain $\Omega_2$ with its local PML region.}
\label{fig_ddm}
\end{figure}

As illustrated in Fig. \ref{fig_ddm}, we use the example of dividing the global computational domain $\Omega$ into two subdomains, $\Omega_1$ and $\Omega_2$, to explain the details of the proposed PML-based source transfer approach. The interior regions of the two subdomains are non-overlapping, and their union covers the entire interior of the global domain. However, as shown in Fig. \ref{fig_ddm}(b) and (c), their respective PML layers extend into each other’s interior regions, creating overlapping zones at the subdomain interfaces. These overlapping PML regions are the core of our method—they serve as the medium through which the equivalent sources are transferred between adjacent subdomains.

\begin{figure}[htbp]
\centering
    \includegraphics[width=\linewidth]{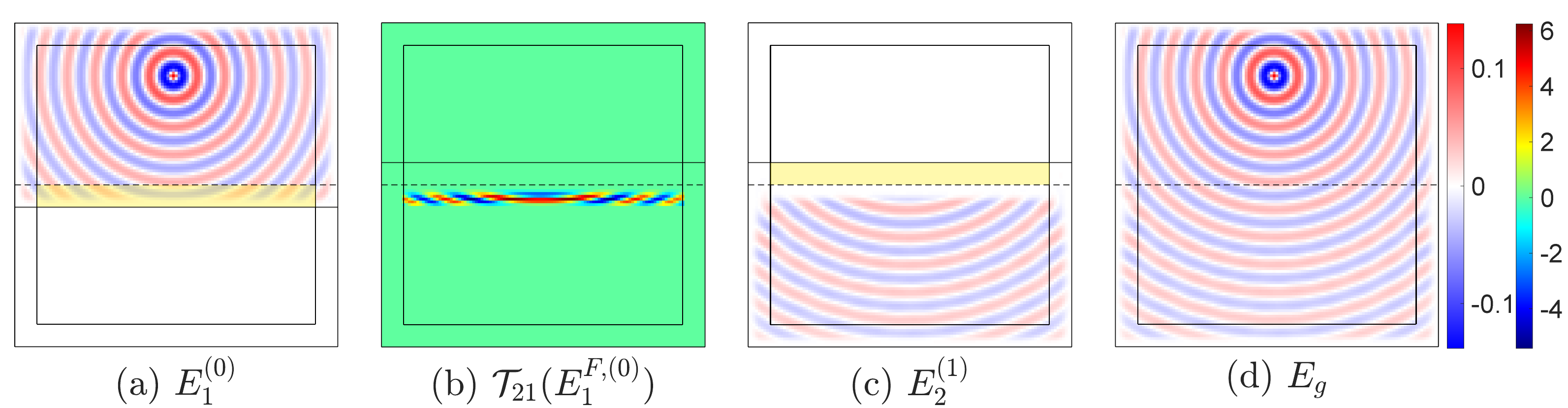}
\caption{The procedure of DDM using a point source modeling. (a) The field $E_1^{(0)}$ of the initial step. (b) The source of $\Omega_2$ using transfer operator $\mathcal{T}_{21}(E_1^{F,(0)})$. (c) The fields $E_2^{(1)}$ of $\Omega_2$ using the transferred source. (d) The combined global field $E_g$.}
\label{fig_point_source}
\end{figure}

We illustrate the procedure using a representative case where a point source is located in subdomain $\Omega_1$ , as depicted in Fig. \ref{fig_point_source}, with the interface between the two subdomains indicated by a dashed line. In the initial step, each subdomain independently solves its local system using only its own source and absorbs outgoing waves through its PML layer, as described by Eqs. (\ref{equ_sys}) and (\ref{equ_pml}):
\begin{equation}
    \nabla_s^2  E_i^{(0)}  + k_0^2 \varepsilon_{r,i} E_i^{(0)} = j\omega\mu_0 J_i, \quad \text{in } \Omega_i,\quad i=1,2,
\end{equation}
where $\nabla_s = \left( \frac{1}{s_x} \partial_x,\ \frac{1}{s_y} \partial_y \right)$ denotes the anisotropic Laplacian under complex coordinate stretching.  In this example, $J_1$ corresponds to a point source excitation, while $J_2=0$. The resulting field $E_1^{(0)}$ is shown in Fig. \ref{fig_point_source}(a). 

To facilitate source transfer, we partition the unknowns in each subdomain $\Omega_i$ into two parts:
\begin{equation}
    E_i = \begin{pmatrix} E_i^I \\ E_i^F \end{pmatrix},
\end{equation}
where $E_i^F$ denotes the electric field within the overlapping PML regions of $\Omega_i$, while $E_i^I$ comprises all remaining unknowns in $\Omega_i$
, including both the physical interior and non-overlapping PML regions. For instance, the yellow area in Fig. \ref{fig_point_source}(a) highlights the overlapping region associated with $E_1^{F}$, which will later act as a transferred source to subdomain $\Omega_2$. 

Based on Maxwell's equations for the 2D TM polarization, the equivalent electric current density of neighbor subdomain $\Omega_p$ is given by:
\begin{equation}
    J_{eq,p} =\frac{1}{j\omega\mu_0} \left(\nabla^2  E_i^{F}  + k_0^2 \varepsilon_{r,p} E_i^{F}\right),\quad \text{in } \Omega_p,
    \label{eq_Jeq}
\end{equation}
which serves as an effective source term for updating the solution in the adjacent subdomain. Notably, the equivalent source in Eq. (\ref{eq_Jeq}) is defined using the standard Helmholtz operator, since from the perspective of $\Omega_p$, $E_i^F$ resides in its physical interior. For convenience, we define a transfer operator as:
\begin{equation}
    \mathcal{T}_{pi}(E_i^{F}) =(\nabla^2 + k_0^2 \varepsilon_{r,p}) E_i^{F}, 
\end{equation}
where the subscript $pi$ indicates that the field $E_i^F$ from subdomain $\Omega_i$ is used to compute a source term in $\Omega_p$. Note that the transfer operator yields a non-zero result only within the overlapping region, as visualized in Fig. \ref{fig_point_source}(b).

The iterative update scheme proceeds as follows: at the $n$-th step, each subdomain solves its local equation with the transferred source term from its neighbor:
\begin{equation}
    \nabla_s^2  E_1^{(n+1)}  + k_0^2 \varepsilon_{r,1} E_1^{(n+1)} = \mathcal{T}_{12}(E_2^{F,(n)}), \quad \text{in } \Omega_1, 
\end{equation}
\begin{equation}
    \nabla_s^2  E_2^{(n+1)}  + k_0^2 \varepsilon_{r,2} E_2^{(n+1)} = \mathcal{T}_{21}(E_1^{F,(n)}), \quad \text{in } \Omega_2, 
\end{equation}
Fig. \ref{fig_point_source}(c) shows $E_2^{(1)}$, the first iteration fields in subdomain $\Omega_2$. In this simple one-way propagation case, the field $E_2^{F,(1)}$ in the overlapping region is nearly zero shown in the yellow area of Fig. \ref{fig_point_source}(c), indicating minimal energy reflection or backward coupling. This observation suggests rapid convergence of the iterative process. The iteration is terminated when the residual energy in the overlapping regions satisfies: 
\begin{equation}
     {\textstyle \sum_{i=1,2}}\left \| E_i^{F,(n)} \right \|<\text{TOL}
\end{equation}
where $\left \|\cdot \right \|$ denotes a suitable norm (e.g., 2-norm), and TOL is a predefined convergence threshold.

Finally, the global field solution is synthesized by summing all contributions from each subdomain over all iterations:
\begin{equation}
     E_g={\textstyle \sum_{n}} (E_1^{(n)}+E_2^{(n)}). 
\end{equation}

As shown in Fig. \ref{fig_point_source}(d), the resulting global solution $E_g$ exhibits a seamless reconstruction of the total field. Notably, in the overlapping regions, the fields are naturally composed of the contributions from both subdomains (e.g., $E_1^{F,(0)} + E_2^{I,(1)}$), yet no visible discontinuity or artifact is observed, confirming the accuracy and smoothness of the decomposition-based computation. This result verifies the effectiveness of the proposed source-transfer scheme using overlapping PMLs in achieving high-fidelity domain decomposition.

To simplify the presentation and clarify the iterative mechanism, the representative example shown in Fig. \ref{fig_point_source} only places the source in subdomain $\Omega_1$, which leads to an alternating-update form of the iteration. However, it is important to emphasize that the proposed method is fully applicable to general scenarios where multiple subdomains contain internal sources. More complex examples involving multiple active subdomains will be presented in the next section to demonstrate this capability.

\subsection{Multi-frontal precondition}
In the proposed domain decomposition framework, the system matrix within each subdomain remains fixed throughout the iterative solution process, while only the right-hand side vector (i.e., the equivalent source) is updated in each iteration. This observation naturally leads to the application of the multi-frontal preconditioner, which efficiently reduces computational overhead of repeated solving linear systems in the iteration by replacing direct matrix inversion with a sequence of sparse triangular factorizations and permutations.

The multi-frontal method is a sparse direct solver based on a divide-and-conquer strategy. It organizes the LU factorization into a sequence of partial factorizations on smaller dense frontal matrices, which are assembled and eliminated recursively along an elimination tree. In the numerical implementation, the multi-frontal precondition is a sparse direct solver that factors the system matrix $\mathbf{A}$ into four components:
\begin{equation}
    \mathbf{A}=\mathbf{P}^T\mathbf{L}\mathbf{U}\mathbf{Q}^T,
\end{equation}
where $\mathbf{P}$ and $\mathbf{Q}$ are permutation matrices to reduce fill-in, and $\mathbf{L}$ and $\mathbf{U}$ are sparse lower- and upper-triangular matrices,respectively.

Once factorized, the system as Eq. (\ref{equ_sys}) can be solved efficiently via forward and backward substitution:
\begin{equation}
    \mathbf{e} = \mathbf{Q} \left( \mathbf{U}^{-1} \left( \mathbf{L}^{-1} \left( \mathbf{P} \mathbf{b} \right) \right) \right).
\end{equation}

This formulation converts the expensive matrix inversion into a chain of matrix-vector multiplications, which can be efficiently executed. For a 2D FDFD discretization with $N$ unknowns, direct sparse solvers (e.g., sparse LU) generally incur a computational complexity of $\mathcal{O}(N^{1.5})$ \cite{Lipton1979}. The multi-frontal method shares the same complexity in its initial factorization step. However, once the matrix is factorized, each subsequent triangular solve (per iteration) has a significantly lower cost of only $\mathcal{O}(N \log N)$ \cite{Wang2016}, as the sparse triangular factors contain only $\mathcal{O}(N \log N)$ non-zero entries, reducing forward/backward substitutions to efficient sparse matrix–vector multiplications. This efficiency gain becomes particularly advantageous in our framework, where multiple iterations are needed to reach convergence, and the matrix remains unchanged across iterations.

Despite its computational advantages, a well-known limitation of the multi-frontal method is its high memory consumption, primarily caused by the fill-in implementation during matrix factorization—particularly in large-scale systems. Fortunately, this issue is effectively alleviated by our domain decomposition approach. By partitioning the global domain into smaller subdomains, the size of each local matrix $\mathbf{A}_i$ is considerably reduced, and the corresponding multi-frontal precondition matrices can be stored with much lower memory cost. Therefore, the combination of domain decomposition and the multi-frontal method enables efficient handling of large-scale problems with manageable memory usage, while preserving the fast convergence and numerical accuracy.

The complete solution process combining multi-frontal precondition and domain decomposition is summarized in Algorithm 1.
\begin{algorithm}[ht]
    \caption{Domain decomposition with $N$ subdomains using multi-frontal precondition.}
\begin{algorithmic}[1]
\STATE \textbf{Define:}\\ 
\STATE $N$: number of subdomains.
\STATE $TOL$: a prescribed tolerance.
\vspace{0.5em}
\STATE \textbf{1. Preprocessing}\\
\FOR{$i = 1$ to $N$}
\STATE Assemble local system matrix $\mathbf{A}_i$ and source term $\mathbf{b}_i$ for subdomain $\Omega_i$.
\STATE Compute multi-frontal precondition matrices: $\mathbf{P}_i$, $\mathbf{L}_i$, $\mathbf{U}_i$ and $\mathbf{Q}_i$.
\ENDFOR
\vspace{0.5em}
\STATE \textbf{2. Iterative Solution}\\
\FOR{$i = 1$ to $N$}
\STATE Initialize local field: $\mathbf{e}_i^{(0)} = \mathbf{Q}_i \left( \mathbf{U}_i^{-1} \left( \mathbf{L}_i^{-1} \left( \mathbf{P}_i \mathbf{b}_i \right)\right)\right)$
\ENDFOR
\WHILE{$ {\sum_{i}}\left \| \mathbf{e}_{i}^{F,(n)} \right \|<\text{TOL}$}
\FOR{$i=1$ to $N$} 
\STATE{Exchange updated fields $\mathbf{e}_{j}^{F,(n-1)}$ from neighboring subdomains $\Omega_j$}
\STATE{Compute total source in subdomain $\Omega_i$: $\mathbf{t}_i^{(n)}=\sum_{j}\mathcal{T}_{ij}(\mathbf{e}_{j}^{F,(n-1)})$}
\STATE{Update solution: $\mathbf{e}_i^{(n)}=\mathbf{Q}_i(\mathbf{U}_i^{-1}(\mathbf{L}_i^{-1}( \mathbf{P}_i\mathbf{t}_i^{(n)})))$}
\ENDFOR
\STATE{$n\gets n+1$}
\ENDWHILE
\vspace{0.5em}
\STATE \textbf{3. Postprocessing}\\
\STATE{Collect the final global field: $\mathbf{e}_g=\sum_{i=1}^{N}{\textstyle \sum_{n}} \mathbf{e}_i^{(n)}$}
\end{algorithmic}
\label{alg1}
\end{algorithm}

\section{Results and discussions}
In this section, we illustrate the validity, efficiency and scalability of the proposed method through two representative examples: the scattering of a dielectric cylinder and a large-scale trapezoidal-shaped topological optical device. All examples are run on our local server, equipped with 2.60 GHz Intel Xeon Gold 6240 processor (64 cores) and 1.4 TB memory.

\subsection{Scattering of a dielectric cylinder}
\begin{figure}[htbp]
\centering
    \includegraphics[width=\linewidth]{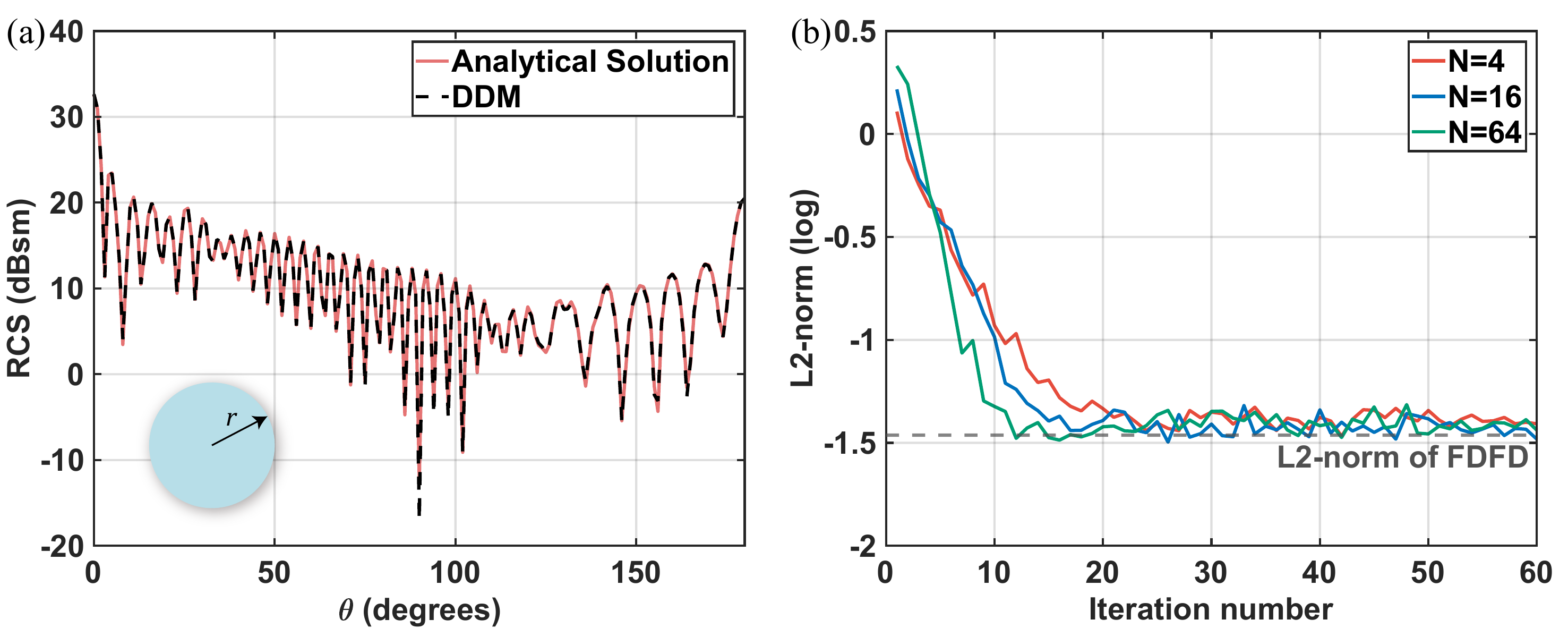}
\caption{(a) RCS of a dielectric cylinder calculated by the analytical solution and the DDM; (b) The relative L2-norm error versus iteration number for different subdomain counts.}
\label{fig_example1}
\end{figure}

We consider the scattering of a 2D dielectric cylinder under TM polarization to verify the accuracy of the proposed method. The cylinder has a radius $r=9\lambda_0$ and a relative permittivity of $\varepsilon_r=4$. The incident wave frequency is 0.3 GHz, and the computational domain is discretized uniformly with 40 points per wavelength, resulting in approximately 36 million unknowns in total. The radar cross section (RCS) computed using the DDM with 64 equally partitioned subdomains (black dashed line) is compared against the analytical Mie series solution \cite{Kong1986} (red solid line), as shown in Fig. \ref{fig_example1}(a). Excellent agreement is observed between the two curves, with a relative L2-norm error of only 3.68\%, demonstrating the accuracy of the proposed method.

Fig. \ref{fig_example1}(b) shows the convergence behavior of the DDM in terms of the relative L2-norm error versus iteration number with respect to the full-domain FDFD solution. Three configurations with different numbers of subdomains are tested: $N=4$, $N=16$ and $N=64$. The dashed horizontal line indicates the L2-norm level of the reference FDFD solution. All cases eventually converge to a comparable accuracy level with the full-domain FDFD result, confirming the robustness and stability of the DDM. Moreover, increasing the number of subdomains significantly accelerates convergence. This improvement can be attributed to enhanced field exchange efficiency across subdomain interfaces. As the number of subdomains increases, a larger number of equivalent sources at internal boundaries are updated independently. This results in finer angular and directional equivalent source transfer, effectively eliminating local field mismatches. This behavior highlights the scalability of the proposed method and suggests that finer domain partitioning can accelerate convergence without sacrificing solution accuracy.

 \begin{table}[htbp]
\centering
\caption{ Computational statistics of DDM with varying number of subdomains.}
\begin{tabular}{@{}ccccccccc@{}}
\toprule
Number of subdomains & 2 & 4 & 8 & 16 & 32 & 64 \\
\midrule
Unknowns per subdomain (million) & 1800 & 900 & 450 & 225 & 112 & 56 \\
Iterations & 32 & 28 & 24 & 22 & 20 & 16 \\
Time (s) & 1372 & 495 & 218 & 101 & 52 & 40 \\
Memory (GB) & 229 & 180 & 150 & 122 & 128 & 136 \\
\bottomrule
\end{tabular}
\label{tab:ddm_performance}
\end{table}

Table \ref{tab:ddm_performance} summarizes the computational performance of the DDM as the number of subdomains increases. For reference, the conventional full-domain FDFD solution takes 958 seconds and consumes 102 GB of memory.

With 16 subdomains, the DDM achieves a runtime of only 101 seconds, corresponding to a speedup factor of approximately 9.5. At $N=64$, the runtime further reduces to 40 seconds, yielding a speedup of 24× relative to the FDFD baseline. This is primarily attributed to the reduced problem size per subdomain and improved parallel efficiency, along with a decrease in the number of iterations required for convergence—which aligns with the analysis in Fig. \ref{fig_example1}(b). However, beyond 32 subdomains, the speedup approaches saturation. This is mainly due to the growing communication overhead and synchronization costs, which offsets the benefit of finer partitioning.

In terms of memory consumption, DDM requires slightly more memory than FDFD. However, this slight increase is well justified by the significant reduction in computational time, making the trade-off both acceptable and worthwhile for large-scale simulations. The increased memory usage primarily stems from the storage of multi-frontal precondition matrices. When the number of subdomains is small, each local matrix remains large, leading to larger fill-in during factorization. As the number of subdomains increases, the size of each local matrix decreases, reducing fill-in and lowering the memory requirement. This trend results in a minimum memory usage of 122 GB at $N=16$. However, when the subdomain count becomes very large (e.g., $N=64$), the memory usage slightly increases again. This is due to two factors: the increased number of precondition matrices that must be stored simultaneously, and the additional communication overhead. Nevertheless, the overall memory remains within a practical range and does not hinder the scalability or efficiency of the method.

\subsection{Trapezoidal-shaped topological photonic crystal waveguide}
Topological photonic crystal (PhC) waveguides have attracted increasing attention in recent years owning to their inherent robustness against structural imperfections and fabrication disorder. However, their complex geometries and the demand for high numerical accuracy pose significant challenges for large-scale electromagnetic simulations. To evaluate the performance of our proposed DDM in such scenarios, we consider a trapezoidal-shaped topological PhC waveguide\cite{Chen2020a} and compare the results with the commercial software COMSOL.

\begin{figure}[htbp]
\centering
    \includegraphics[width=\linewidth]{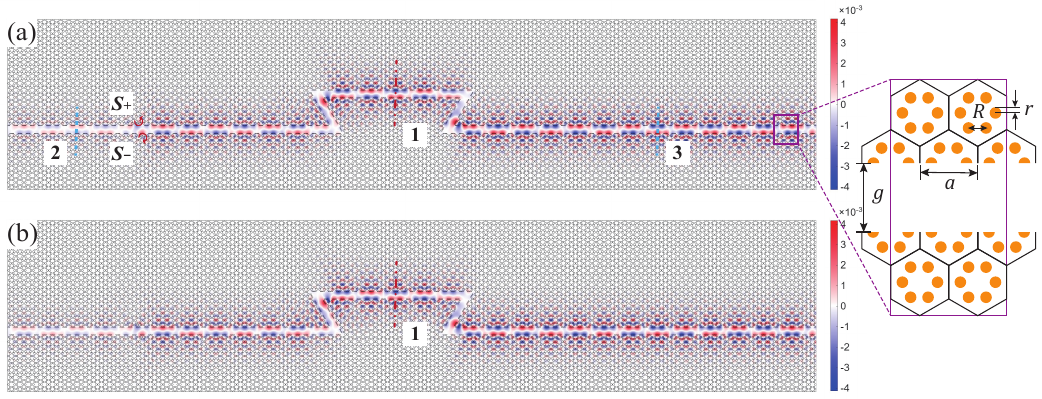}
\caption{Simulated electric field distribution and geometric structure of the topological PhC waveguide; (a) Simulation results calculated by DDM. (b) Simulation results calculated by COMSOL.}
\label{fig_example2}
\end{figure}

The geometry of the photonic crystal and the corresponding line-defect waveguide are illustrated in the right of Fig. \ref{fig_example2}. Each unit cluster consists of six dielectric cylinders placed at the vertices of a hexagon with a side length of $R=6$ mm, and the lattice constant is set to $a=2.8R$. The cylinders have a relative permittivity $\varepsilon_r=11.7$ and radius $r=2$ mm, while the defect gap width is $g=21$ mm. The working frequency is fixed at 7.96 GHz. Reverse orbital angular momentum (OAM) sources are applied at the upper and lower edges of the line defect, each realized by a square excitation positioned inside a hexagon near the edge. The overall structure size is approximately $95\lambda_0 \times 20\lambda_0$ ($\lambda_0$ is the free-space wavelength), discretized with 85 grids per $\lambda_0$ (considering high dielectric contrast, about 25 grids per wavelength inside the dielectric).  So, the total number of unknowns is 13,727,500. The entire computational domain is partitioned into 64 subdomains for the DDM simulation.

\begin{figure}[htbp]
\centering
    \includegraphics[width=0.8\linewidth]{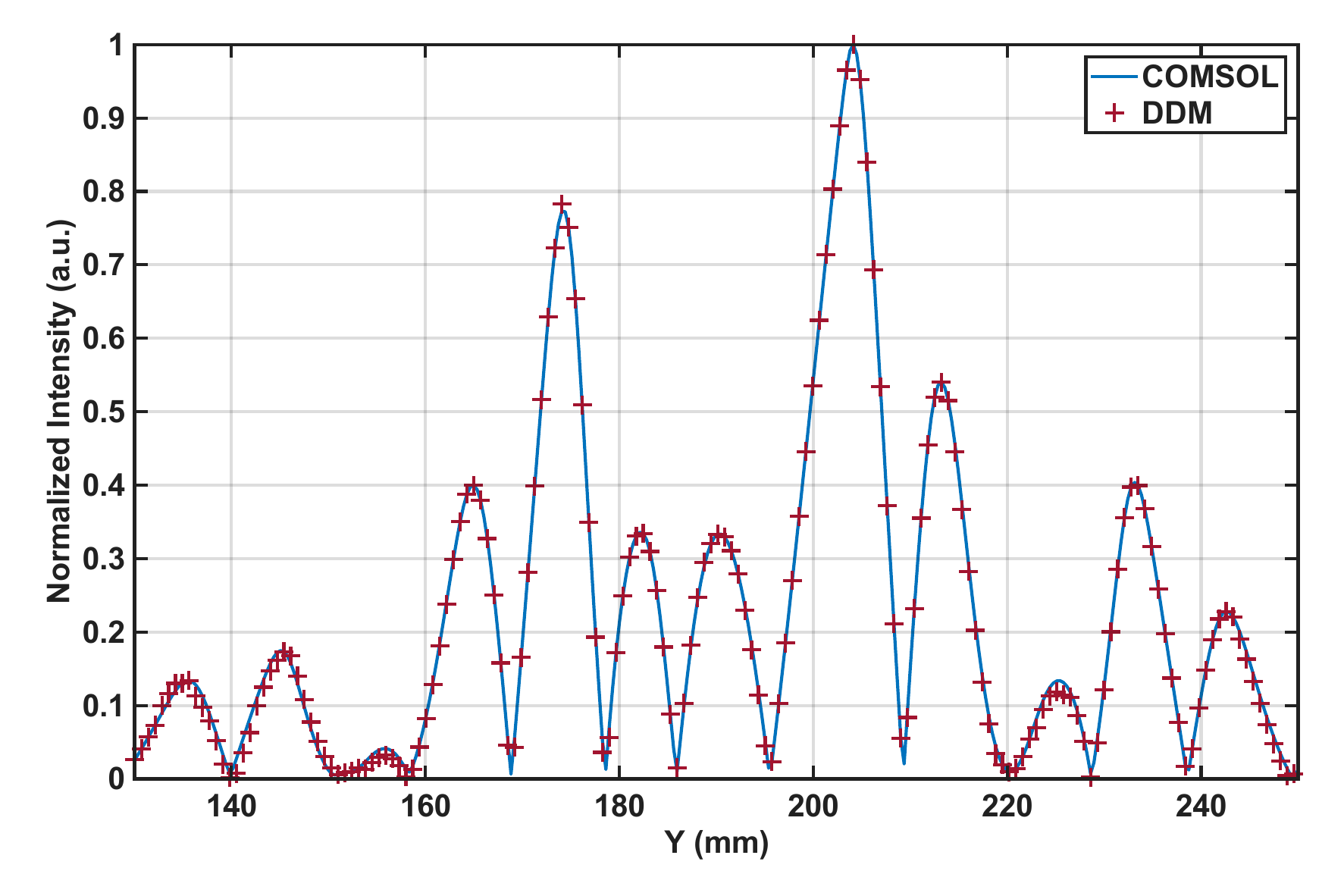}
\caption{Normalized electric field alone plane 1 in Fig. \ref{fig_example2} by the proposed algorithm and COMSOL software.}
\label{fig_Eline}
\end{figure}

In Fig. \ref{fig_example2}, the distribution of the electric-field amplitude computed by the proposed DDM is shown, along with the results simulated by COMSOL. An excellent agreement between the two methods can be observed, indicating that our solver is capable of capturing the topological edge-state propagation accurately. To further validate the accuracy of our solver, Fig. \ref{fig_Eline} compares the normalized electric field amplitude along the reference plane 1 indicated in Fig. \ref{fig_example2}. The results obtained from DDM and COMSOL almost perfectly overlap, confirming that our approach not only reproduces the global field profile but also yields accurate field values along critical waveguide sections.

The blue dashed lines in Fig. \ref{fig_example2}(a) represent the planes used for evaluating transmitted and reflected power through $U=1/2\int_{l}\mathrm{Re}(\mathbf{E}\times\mathbf{H}^*)$. Then, we define the backscattering ratio as $U_2/(U_2+U_3)$ and it is calculated to be 7.9\%, confirming the robustness of the topological channel even in the presence of half-cell truncation and sharp bends. Furthermore, the strong confinement of the field within the defect channel confirms that radiation leakage into the surrounding bulk photonic crystal is effectively suppressed.

 Furthermore, Table \ref{tab:example2} summarizes the computational performance of the different methods. Since COMSOL software is based on FEM, the comparison is carried out under comparable accuracy for fairness. The relative L2-norm errors of FDFD and DDM with respect to the COMSOL reference are 2.62\% and 2.87\%, respectively. In terms of computation time, DDM significantly outperforms the other two methods, completing the computation within 24 s, which corresponds to a speedup factor of approximately 13.4 compared with COMSOL and 18.3 compared with conventional FDFD. Regarding memory consumption, DDM requires 40 GB, which is moderately higher than COMSOL (23 GB) and FDFD (32 GB). The additional cost of DDM mainly originates from storing multi-frontal factors and communication overhead, whereas COMSOL benefits from highly optimized internal memory management. Nevertheless, considering the substantial acceleration achieved by DDM, this increase in memory demand is acceptable and does not compromise overall efficiency.
\begin{table}[htbp]
\centering
\caption{Computational statistics of the topological PhC waveguide.}
\begin{tabular}{lccc}
\toprule
 & \textbf{COMSOL} & \textbf{FDFD} & \textbf{DDM} \\
\midrule
Computation time (s) & 321 & 440 & 24 \\
Memory usage (GB) & 23 & 32 & 40 \\
Relative error & -- & 2.62\% & 2.87\% \\
\bottomrule
\end{tabular}
\label{tab:example2}
\end{table}

\section{Conclusion}
In this work, we have developed a parallel overlapping DDM for accelerating the calculation for large-scale complex nanostructures. By utilizing PML regions as overlapping zones for source transfer, the approach ensures consistent wave behavior, preserves physical continuity of electromagnetic fields, and enables efficient parallelization with minimal communication overhead. The incorporation of multi-frontal preconditioning and OpenMP parallelization significantly achieves both numerical accuracy and computational efficiency. Compared with full-domain FDFD and commercial COMSOL software in numerical examples, the proposed approach demonstrates substantial speedup—up to more than an order of magnitude—without compromising accuracy, at the cost of only modestly increased memory consumption. These results highlight the method’s potential for addressing computational challenges in next-generation photonic integrated circuits, VR/AR devices, and other large-area electromagnetic systems. Future work will extend this framework to 3D problems and explore advanced load-balancing strategies for massively parallel computing platforms.

\begin{backmatter}

\bmsection{Disclosures}
The authors declare no conflicts of interest.

\bmsection{Data availability} Data underlying the results presented in this paper are not publicly available at this time but may be obtained from the authors upon reasonable request.

\end{backmatter}

\bibliography{References.bib}

\begin{thebibliography}{10}
\newcommand{\enquote}[1]{``#1''}

\bibitem{He2022}
D.~Yao, P.~H. He, H.~C. Zhang, \emph{et~al.}, \enquote{Miniaturized photonic and microwave integrated circuits based on surface plasmon polaritons,} {\protect\JournalTitle{Progress In Electromagnetics Research}} \textbf{175}, 105--125 (2022).

\bibitem{Ning2024}
S.~Ning, H.~Zhu, C.~Feng, \emph{et~al.}, \enquote{Photonic-electronic integrated circuits for high-performance computing and {AI} accelerators,} {\protect\JournalTitle{Journal of Lightwave Technology}} \textbf{42}, 7834--7859 (2024).

\bibitem{Huang2024}
C.~Huang and W.~E.~I. Sha, \enquote{A parallel block preconditioner-based {VIE-FFT} algorithm for modeling the electromagnetic response from nanostructures,} {\protect\JournalTitle{IEEE Transactions on Antennas and Propagation}} \textbf{72}, 1051--1056 (2024).

\bibitem{Malevich2025}
Y.~Malevich, M.~S. Ergoktas, G.~Bakan, \emph{et~al.}, \enquote{Very-large-scale reconfigurable intelligent surfaces for dynamic control of terahertz and millimeter waves,} {\protect\JournalTitle{Nature Communications}} \textbf{16}, 2907 (2025).

\bibitem{Taygur2019}
M.~M. Taygur, I.~O. Sukharevsky, and T.~F. Eibert, \enquote{Investigation of massive {MIMO} scenarios involving rooftop propagation by bidirectional ray-tracing,} {\protect\JournalTitle{Progress In Electromagnetics Research C}} \textbf{91}, 129--142 (2019).

\bibitem{Sullivan2013}
D.~M. Sullivan, \emph{Electromagnetic simulation using the FDTD method} (John Wiley \& Sons, 2013).

\bibitem{Jin2015}
J.-M. Jin, \emph{The finite element method in electromagnetics} (John Wiley \& Sons, 2015).

\bibitem{Minkov2020}
M.~Minkov, I.~A.~D. Williamson, L.~C. Andreani, \emph{et~al.}, \enquote{Inverse design of photonic crystals through automatic differentiation,} {\protect\JournalTitle{ACS Photonics}} \textbf{7}, 1729--1741 (2020).

\bibitem{Mukawa2009}
H.~Mukawa, K.~Akutsu, I.~Matsumura, \emph{et~al.}, \enquote{A full-color eyewear display using planar waveguides with reflection volume holograms,} {\protect\JournalTitle{Journal of The Society for Information Display - J SOC INF DISP}} \textbf{17}, 185--193 (2009).

\bibitem{Kramer2023}
E.~L. Kramer, Q.~Chao, M.~Schaub, \emph{et~al.}, \enquote{End-to-end sensor system modeling and validation for {AR/VR/MR} applications,} in \emph{Physics and Simulation of Optoelectronic Devices XXXI,}  vol. 12415 (SPIE, 2023), pp. 182--193.

\bibitem{Low2022}
M.~J. Low, T.~M. Rohith, B.~Kim, \emph{et~al.}, \enquote{Refractive-diffractive hybrid optics array: Comparative analysis of simulation and experiments,} {\protect\JournalTitle{Journal of Optics}} \textbf{24}, 055401 (2022).

\bibitem{Arbabi2016}
A.~Arbabi, E.~Arbabi, S.~M. Kamali, \emph{et~al.}, \enquote{Miniature optical planar camera based on a wide-angle metasurface doublet corrected for monochromatic aberrations,} {\protect\JournalTitle{Nature Communications}} \textbf{7}, 13682 (2016).

\bibitem{Su2018}
Y.~Su and Z.~N. Chen, \enquote{A flat dual-polarized transformation-optics beamscanning luneburg lens antenna using {PCB}-stacked gradient index metamaterials,} {\protect\JournalTitle{IEEE Transactions on Antennas and Propagation}} \textbf{66}, 5088--5097 (2018).

\bibitem{Ding2022}
F.~Ding, \enquote{A review of multifunctional optical gap-surface plasmon metasurfaces,} {\protect\JournalTitle{Progress In Electromagnetics Research}} \textbf{174}, 55--73 (2022).

\bibitem{Pestourie2018}
R.~Pestourie, C.~{P{\'e}rez-Arancibia}, Z.~Lin, \emph{et~al.}, \enquote{Inverse design of large-area metasurfaces,} {\protect\JournalTitle{Optics Express}} \textbf{26}, 33732--33747 (2018).

\bibitem{Skarda2022}
J.~Skarda, R.~Trivedi, L.~Su, \emph{et~al.}, \enquote{Low-overhead distribution strategy for simulation and optimization of large-area metasurfaces,} {\protect\JournalTitle{npj Computational Materials}} \textbf{8}, 78 (2022).

\bibitem{Zu2025}
X.~Zu, X.~Sun, W.~Yan, \emph{et~al.}, \enquote{Fast and efficient inverse design framework for multifunctional metalenses,} {\protect\JournalTitle{Laser \& Photonics Reviews}} \textbf{19}, 2400886 (2025).

\bibitem{Li2022}
Z.~Li, R.~Pestourie, J.-S. Park, \emph{et~al.}, \enquote{Inverse design enables large-scale high-performance meta-optics reshaping virtual reality,} {\protect\JournalTitle{Nature Communications}} \textbf{13}, 2409 (2022).

\bibitem{Xiao2025}
F.~Xiao, J.~Wang, Z.~Xiong, and Y.~Chen, \enquote{Numerical approximation of slowly varying envelope in finite element electromagnetism: A ray-wave method of modeling multi-scale devices,} {\protect\JournalTitle{Optics Express}} \textbf{33}, 12603--12614 (2025).

\bibitem{Wyrowski2011}
F.~Wyrowski and M.~Kuhn, \enquote{Introduction to field tracing,} {\protect\JournalTitle{Journal of Modern Optics}} \textbf{58}, 449--466 (2011).

\bibitem{Wang2024}
J.~Wang, Z.~Wang, L.~Liu, and Y.~Chen, \enquote{Multi-scale multi-domain hybrid finite element modeling of light propagation,} {\protect\JournalTitle{Electromagnetic Science}} \textbf{2}, 1--10 (2024).

\bibitem{Lokar2019a}
Z.~Lokar, B.~Lipovsek, A.~Razzaq, \emph{et~al.}, \enquote{Coupled modelling approach for optimization of bifacial silicon heterojunction solar cells with multi-scale interface textures,} {\protect\JournalTitle{Optics Express}} \textbf{27}, A1554--A1568 (2019).

\bibitem{Lee2005}
S.-C. Lee, M.~N. Vouvakis, and J.-F. Lee, \enquote{A non-overlapping domain decomposition method with non-matching grids for modeling large finite antenna arrays,} {\protect\JournalTitle{Journal of Computational Physics}} \textbf{203}, 1--21 (2005).

\bibitem{Vouvakis2006}
M.~Vouvakis, Z.~Cendes, and J.-F. Lee, \enquote{A {FEM} domain decomposition method for photonic and electromagnetic band gap structures,} {\protect\JournalTitle{IEEE Transactions on Antennas and Propagation}} \textbf{54}, 721--733 (2006).

\bibitem{Xue2012}
M.-F. Xue and J.-M. Jin, \enquote{Nonconformal {FETI-DP} methods for large-scale electromagnetic simulation,} {\protect\JournalTitle{IEEE Transactions on Antennas and Propagation}} \textbf{60}, 4291--4305 (2012).

\bibitem{Gao2016}
H.-W. Gao, M.-L. Yang, and X.-Q. Sheng, \enquote{Nonconformal {FETI-DP} domain decomposition methods for {FE-BI-MLFMA},} {\protect\JournalTitle{IEEE Transactions on Antennas and Propagation}} \textbf{64}, 3521--3532 (2016).

\bibitem{Peng2011}
Z.~Peng and J.-F. Lee, \enquote{Non-conformal domain decomposition method with mixed true second order transmission condition for solving large finite antenna arrays,} {\protect\JournalTitle{IEEE Transactions on Antennas and Propagation}} \textbf{59}, 1638--1651 (2011).

\bibitem{Jia2019}
P.-H. Jia, L.~Lei, J.~Hu, \emph{et~al.}, \enquote{Twofold domain decomposition method for the analysis of multiscale composite structures,} {\protect\JournalTitle{IEEE Transactions on Antennas and Propagation}} \textbf{67}, 6090--6103 (2019).

\bibitem{Zhao2016}
R.~Zhao, J.~Hu, H.~Zhao, \emph{et~al.}, \enquote{Solving {EM} scattering from multiscale coated objects with integral equation domain decomposition method,} {\protect\JournalTitle{IEEE Antennas and Wireless Propagation Letters}} \textbf{15}, 742--745 (2016).

\bibitem{Zhao2020}
R.~Zhao, Y.~Chen, X.-M. Gu, \emph{et~al.}, \enquote{A local coupling multitrace domain decomposition method for electromagnetic scattering from multilayered dielectric objects,} {\protect\JournalTitle{IEEE Transactions on Antennas and Propagation}} \textbf{68}, 7099--7108 (2020).

\bibitem{Wang2017}
W.-J. Wang, R.~Xu, H.-Y. Li, \emph{et~al.}, \enquote{Massively parallel simulation of large-scale electromagnetic problems using one high-performance computing scheme and domain decomposition method,} {\protect\JournalTitle{IEEE Transactions on Electromagnetic Compatibility}} \textbf{59}, 1523--1531 (2017).

\bibitem{MacKie2018}
B.~MacKie-Mason, Y.~Shao, A.~Greenwood, and Z.~Peng, \enquote{Supercomputing-enabled first-principles analysis of radio wave propagation in urban environments,} {\protect\JournalTitle{IEEE Transactions on Antennas and Propagation}} \textbf{66}, 6606--6617 (2018).

\bibitem{Chen2013}
Z.~Chen and X.~Xiang, \enquote{A source transfer domain decomposition method for {Helmholtz} equations in unbounded domain,} {\protect\JournalTitle{SIAM Journal on Numerical Analysis}} \textbf{51}, 2331--2356 (2013).

\bibitem{Xiang2013}
Z.~C. Xiang and Xueshuang, \enquote{A source transfer domain decomposition method for {Helmholtz} equations in unbounded domain part ii: Extensions,} {\protect\JournalTitle{Numerical Mathematics: Theory, Methods and Applications}} \textbf{6}, 538--555 (2013).

\bibitem{Leng2015}
L.~Wei, \enquote{A fast propagation method for the {Helmholtz} equation,} {\protect\JournalTitle{Chinese Journal of Engineering Mathematics}} \textbf{32}, 726--742 (2015).

\bibitem{Leng2022}
W.~Leng and L.~Ju, \enquote{Trace transfer-based diagonal sweeping domain decomposition method for the {Helmholtz} equation: Algorithms and convergence analysis,} {\protect\JournalTitle{Journal of Computational Physics}} \textbf{455}, 110980 (2022).

\bibitem{Dai2022}
R.~Dai, A.~Modave, J.-F. Remacle, and C.~Geuzaine, \enquote{Multidirectional sweeping preconditioners with non-overlapping checkerboard domain decomposition for {Helmholtz} problems,} {\protect\JournalTitle{Journal of Computational Physics}} \textbf{453}, 110887 (2022).

\bibitem{Xia2010}
J.~Xia, S.~Chandrasekaran, M.~Gu, and X.~S. Li, \enquote{Superfast multifrontal method for large structured linear systems of equations,} {\protect\JournalTitle{SIAM Journal on Matrix Analysis and Applications}} \textbf{31}, 1382--1411 (2010).

\bibitem{Qiao2011}
P.-f. Qiao, W.~E.~I. Sha, W.~C.~H. Choy, and W.~C. Chew, \enquote{Systematic study of spontaneous emission in a two-dimensional arbitrary inhomogeneous environment,} {\protect\JournalTitle{Physical Review A}} \textbf{83}, 043824 (2011).

\bibitem{Sha2010}
W.~E.~I. Sha, W.~C.~H. Choy, and W.~C. Chew, \enquote{A comprehensive study for the plasmonic thin-film solar cell with periodic structure,} {\protect\JournalTitle{Optics Express}} \textbf{18}, 5993--6007 (2010).

\bibitem{Chew1997a}
W.~C. Chew, J.~M. Jin, and E.~Michielssen, \enquote{Complex coordinate stretching as a generalized absorbing boundary condition,} {\protect\JournalTitle{Microwave and Optical Technology Letters}} \textbf{15}, 363--369 (1997).

\bibitem{Berenger1996}
J.-P. Berenger, \enquote{Three-dimensional perfectly matched layer for the absorption of electromagnetic waves,} {\protect\JournalTitle{Journal of Computational Physics}} \textbf{127}, 363--379 (1996).

\bibitem{Lipton1979}
R.~J. Lipton, D.~J. Rose, and R.~E. Tarjan, \enquote{Generalized nested dissection,} {\protect\JournalTitle{SIAM Journal on Numerical Analysis}} \textbf{16}, 346--358 (1979).

\bibitem{Wang2016}
S.~Wang, X.~S. Li, F.-H. Rouet, \emph{et~al.}, \enquote{A parallel geometric multifrontal solver using hierarchically semiseparable structure,} {\protect\JournalTitle{ACM Trans. Math. Softw.}} \textbf{42}, 1--21 (2016).

\bibitem{Kong1986}
J.~A. Kong, \emph{Electromagnetic Wave Theory} (Wiley, New York, NY, USA, 1986).

\bibitem{Chen2020a}
M.~L.~N. Chen, L.~J. Jiang, Z.~Lan, and W.~E.~I. Sha, \enquote{Pseudospin-polarized topological line defects in dielectric photonic crystals,} {\protect\JournalTitle{IEEE Transactions on Antennas and Propagation}} \textbf{68}, 609--613 (2020).

\end{thebibliography}

\end{document}